\documentclass[12pt,preprint]{aastex}
\usepackage{emulateapj5, epsfig}

\def\asca{{\sl ASCA }}

\def\xte{{\sl RXTE }}
\def\swift{{\sl SWIFT }}

\def\chandra{{\sl Chandra }}
\def\circinus{Circinus~X-1~}
\def\cirx1{Cir~X-1~}
\def\ergsec{\hbox{erg s$^{-1}$ }}
\def\ergcm{\hbox{erg cm$^{-2}$ s$^{-1}$ }}

\def\Msun{$M_{\odot}$ }

\def\it{\sl}

\def\lapp{\ifmmode\stackrel{<}{_{\sim}}\else$\stackrel{<}{_{\sim}}$\fi}

\def\gapp{\ifmmode\stackrel{>}{_{\sim}}\else$\stackrel{>}{_{\sim}}$\fi}
\def\spose#1{\hbox to 0pt{#1\hss}}

\def\approxlt{\mathrel{\spose{\lower 3pt\hbox{$\sim$}}
        \raise 2.0pt\hbox{$<$}}}
\def\approxgt{\mathrel{\spose{\lower 3pt\hbox{$\sim$}}
        \raise 2.0pt\hbox{$>$}}}

\slugcomment{accepted for publication September 4, 2007 by The Astrophysical Journal}

\shorttitle{THE WARM ABSORBER IN CIR~X-1}

\shortauthors{SCHULZ et al.}

\begin{document}

\title{The Variable Warm Absorber in Circinus~X-1}
\author{
N. S. Schulz\altaffilmark{1},
T. E. Kallman\altaffilmark{2},
D. K. Galloway\altaffilmark{3},
and
W. N. Brandt\altaffilmark{4}}
\altaffiltext{1}{Kavli Institute for Astrophyscis and Space Research, Massachusetts Institute of Technology,
Cambridge, MA 02139.}
\altaffiltext{2}{Goddard Space Flight Center, NASA,
Greenbelt, MD .}
\altaffiltext{3}{School of Physics, University of Melbourne,
Parkville 3010, Australia.}
\altaffiltext{4}{Department of Astronomy \& Astrophysics, 525 Davey Laboratory,
The Pennsylvania State University, University Park, PA, 16802.}

\begin{abstract}
We observed \circinus twice during a newly reached low-flux phase near zero orbital phase using
the High-Energy Transmission Grating Spectrometer (HETGS) onboard \chandra.  
In both observations the source did not show the P Cygni lines we observed during
the high-flux phases of the source in 2000 and 2001.
During pre-zero phase the source did not exhibit significant variability and exhibited an 
emission-line spectrum rich in H- and He-like lines from high Z elements such as
Si, S, Ar, and Ca. The light curve in the post-dip observation showed 
quiescent and flaring episodes. Only in these flaring episodes is the source luminosity 
significantly higher than observed during the pre-zero phase.
We analyzed all high resolution X-ray
spectra by fitting photoionization and absorption models from the most recent version of
the XSTAR code. 
The pre-zero phase spectrum could be fully modeled with a
very hot photoionized plasma with an ionization parameter of $log~\xi = 3.0$, down from
$log~\xi = 4.0$ in the high-flux state.  
The ionization balances we measure from the spectra during the post-zero phase
episodes are significantly different. Both 
episodes feature absorbers with variable high columns, ionization parameter, and luminosity.
While cold absorption remains at levels quite similar to the one observed in previous years, 
the new observations show unprecedented levels of variable warm absorption.  
The line emissivities also indicate that the observed low source luminosity is
inconsistent with a static hot accretion disk corona (ADC), an effect that seems common to other 
near edge-on ADC sources as well. We conclude that unless there exists some means
of coronal heating other than X-rays, the true source luminosity is likely much higher
and we observe obscuration in analogy to the
extragalactic Seyfert II sources.  We discuss possible consequences and relate 
cold, luke-warm, warm, and hot absorbers to dynamic accretion scenarios.
\end{abstract}

\keywords{
stars: individual (Cir~X-1) ---
stars: neutron ---
X-rays: stars ---
binaries: close ---
accretion: accretion disks ---
techniques: spectroscopic}

\section{Introduction}

Not many X-ray binaries show as large a variety of X-ray features
and properties as \circinus. Since its discovery~\citep{margon1971}, 
this X-ray binary has appeared at various X-ray brightness levels ranging
from only a few to several hundreds of mCrab~\citep{parkinson2003}.
Its long-term brightness featured flaring activity at increasing
levels during the 1970s and 1980s and a steady increase in overall
brightness during the 1990s until at some point it reached a
flux of about 1.5 Crab. The discovery of broad P~Cygni line profiles
during these bright periods~\citep{brandt2000,schulz2002} demonstrated
the presence of a high-velocity outflow likely in the form of 
an accretion disk wind. The simultanuous presence of possibly relativistic
jets~\citep{fender1998,fender2004a} underlines the highly violent 
nature of the source at the time.

Since then \cirx1 has calmed considerably in its X-ray emission
and although the X-ray light curve still shows its characteristic pattern
with an orbital period of 16.6 days \citep{kaluzienski1976}, its 
outbursting power near zero phase also has diminished to only a fraction
of the levels it showed a few years ago. In recent months the source was
only barely or not detected with the \xte ASM~\citep{jonker2007a}. In most repects the nature of \cirx1 
is still poorly understood. Despite advances
in recent years, there remains uncertainty about even the most
basic properties of this system. Recent observations of quasi-periodic 
oscillations (QPOs) added more incentives to a long list of indirect evidence
that the compact object in \cirx1 is in fact a neutron star 
\citep{shirey1999, tennant1986, qu2001,tauris1999,boutloukos2006}. 
Some new insights on the nature of its companion came from a recent 
analysis by \citet{jonker2007b} who argue for an earlier A to late B-type
star. \citet{moneti1992} had detected a heavily reddened optical counterpart
with strong, asymmetric and variable H$\alpha$ emission 
\citep{whelan1977, mignani1997, johnston2001}.

The system's orientation has become a much debated issue~\citep{schulz2006}. 
Perhaps one of the most exciting but also controversial features of \cirx1 is
the existence of a possibly ultrarelativistic radio jet. Under the assumption
that the jet was launched during the most recent periastron passage, 
\citet{fender2004a} deduce an upper limit to the inclination
of about 11$^{\circ}$. Most recent projections of orbital parameters 
postulate an angle of no less than 14$^{\circ}$ \citep{jonker2007b}. 
Recently \citet{heinz2007} found evidence for a parsec scale X-ray jet from
\cirx1 which also appears consistent with some lower inclination for an unbended jet.
X-ray spectral studies do not
directly provide evidence about inclination since
no eclipses are observed. However, extensive studies with \asca and \xte~\citep{brandt1996,
shirey1999}, and later with \chandra~\citep{brandt2000, schulz2002} produced
arguments for a highly inclined viewing angle including an equatorial
wind-like outflow that now seems to be a quite typical property of these
types of X-ray binaries \citep{blundell2001,miller2006a,miller2006b}.

In this respect \circinus belongs to a now long list of Galactic
microquasars~\citet{mirabel2001}  named in analogy to their extragalactic
cousins, the Seyfert galaxies. One of the trademarks of active galactic
nuclei is, besides an X-ray bright nucleus and prominent relativistic
jets, the existence of warm absorbers. There are various types of
X-ray absorbers in X-ray sources. Widely known in Galactic X-ray binaries
is photoelectric 
absorption due to cool and lowly ionized matter of the interstellar
medium~\citep{morrisson1983,wilms2000,paerels2001,juett2004,juett2006}
as well as circumstellar and circumdisk material~\citep{watanabe2003, schulz2002a, juett2006}.
The most prominent indication for the existence of large amount of such cool
circumstellar material is the existence of significant line fluorescence
as well orbitally correlated column densities. The existence of Seyfert-like warm
absorber outflows in microquasars other than \cirx1 has recently also been observed
with \chandra in GX 339-4 and GROJ 1655-40~\citep{miller2004,miller2006b}. 
The term warm absorber was introduced during the mid-1980s by
\citet{halpern1984} as an illuminated plasma of moderate 
electron temperature ($<$ 10$^7$ K) which shows similar levels of ionization
as a collisionally ionized hot gas. Quite impressive examples of warm absorbers 
have been shown for AGNs like MCG6-30-15~\citep{lee2002} and NGC 3783~\citep{kaspi2002}.
A recent overview of warm absorber properties in AGN was given by ~\citet{blustin2005}. 

In this third paper, we present results from two new observations of the 
Galactic microquasar \circinus
performed during a phase where the source was at an overall extremely
low flux level. As in papers~I and II (\citealt{brandt2000} (BSI) and 
\citealt{schulz2002} (SBII)) the 
observations were performed with the High-Energy Transmission Grating
Spectrometer (HETGS) onboard \chandra at ``zero phase.'' Zero phase in
Cir~X-1 is thought to be associated with the periastron passage of
the neutron star. The observations include a shorter exposure during the 
pre-zero phase dip and a longer exposure at zero phase where the
source usually exhibits strong variability. The second of these observations
was also the basis for the discovery of the parsec-scale X-ray jet by
\citet{heinz2007}. 
Throughout this paper, we adopt a distance to Cir~X-1 of 6~kpc
\citep{stewart1993, case1998} and an interstellar 
column density of about $2\times10^{22}$ cm$^{-2}$ (e.g., \citealt{predehl1995}). 

\smallskip
\includegraphics[angle=0,width=8.5cm]{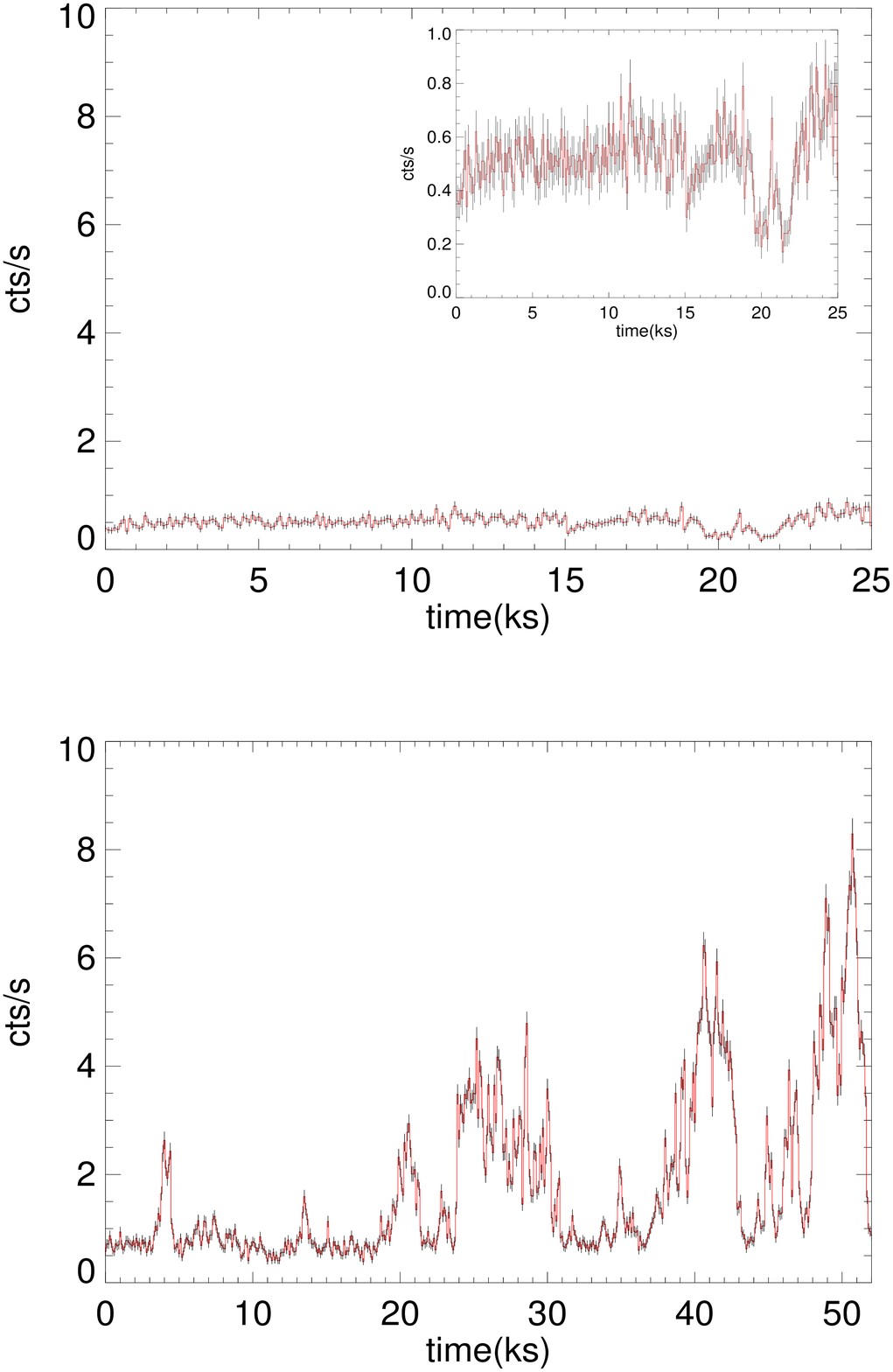}
\figcaption{The light curves of obs.~III (top) and IV (bottom). The data
were binned by adding all HETG 1st order photons from 1.6 to 10~\AA. The y-scale
of the two curves was kept the same for better comparison. The inset in the top
image shows a blow up of III.
\label{lightcurves}}
\smallskip

\section{Chandra Observations}

\cirx1 was observed with the HETGS (see \citealt{canizares2005} for a detailed description)  
on 2005 January~3 (starting at 01:14:52 UT) for 25~ks (observation~III, OBSID 6148) and on 
2005 June~2 (starting at 23:44:51 UT) for another 50~ks 
(observation~IV, OBSID 5478). The denominations `III' and `IV' stem from the fact that we
observed \cirx1 a few years earlier in a very similar manner and 
instrument configuration which were labeled as observations~I and~II (SBII).
Both observations were performed near zero phase of the binary orbit. Observation~III
was performed  during an often occurring pre-zero phase dip at phase 0.96 -- 0.99, while
obs.~IV occurred a few phase bins later at phases 1.06 -- 1.10 
(adopting the ephemeris of \citealt{glass1994}). Both observations were
configured the same way. The detector array (ACIS-S) consists of 6 CCDs which
were clocked choosing a 512 pixel row subarray placed near the readout in order to
reduce the frame time to about 1.7~s. There were two distinct differences in the setup
of the two observations which stemmed from two erroneous settings during obs.~III.
While the first produced a different off-axis angle of the telescope with the
only consequence that chip gaps in the effective area fall into different wavelength
bands, the second was a bit more costly. Due to an incorrectly placed exclusive 
count window filter on the central CCD (and we detect no source photons above
10~\AA), the spectra in obs.~III
possess only about 50$\%$ of the anticipated exposure.

\begin{table*}
\begin{center}
{\sc TABLE~1: CHANDRA HETGS X-RAY OBSERVATIONS IN 2005}
\begin{tabular}{lcccccc}
          &  & & & & & \\
\tableline
 Obsid & label & Start Date & Start Time & Exposure & HEG -1st rate    & Phase\\
       &       &     [UT]        & [UT]       & [ks]     & cts s$^{-1}$ & range\\
\tableline
          & & & & & & \\
 6148 & III & Jan 03 2005 & 01:14:52 &  24.48 & 0.228 & 0.96 -- 0.99 \\
 5478 & IV  & Jun 02 2005 & 23:44:51 &  51.62 & 0.524 & 1.06 -- 1.10 \\
          & & & & & & \\
\tableline
\end{tabular}
\end{center}
\end{table*}

All observations were reprocessed using CIAO3.4 with the most recent
CALDB products. In order to fine tune the wavelength scale we redetermined the zero-order
position in both observations. For obs.~IV we have a strong 0th order point source with
read out trace and positive and negative order dispersion. This allows us to reach
a positional accuracy of about 1/4 of a detector pixel ensuring a wavelength
scale accuracy of about 1/8 of a resolution element, i.e. 0.0025~\AA~for MEG and 
0.0015~\AA~for HEG spectra. In obs.~III the 0th order was filtered for every 10th photon and
weak, and due to the misplaced exclusion window we do not have a read out trace and
significant negative order spectra. This severely limited our ability to determine
a precise 0th order position. Therefore for obs.~III we only reach an accuracy of
0.010~\AA~for MEG and 0.005~\AA~for HEG in our wavelength scale.
For transmission gratings the dispersion scale is linear in
wavelength and we perform the spectral analysis always in wavelength space to avoid non-linear
binning.
We used standard wavelength redistribution matrix files (RMF)
but generated ancillary response files (ARFs) using the provided aspect solutions, bad pixel maps, and 
CCD window filters.
\footnote{see \url{http://asc.harvard.edu/ciao/threads/}}
For all the observations we
generated spectra and analysis products for the 1st orders only.

Figure~\ref{lightcurves} shows the light curves of the two observations. Note that since obs.~III
(top) occurred at pre-zero orbital phase, the count rate appears exceptionally low. However, even
compared with observation~II
from the year 2000 at similar phase, the rate has dropped by over two orders of magnitude.
Observation~IV (bottom) shows not only the expected higher rate, but also quite significant variability. Besides
episodes of relative quiescence the source now engages in frequent flares. 
Though this is not unexpected during this orbital phase, the observed activity has 
a somewhat different appearance compared to the activity in previous years where the flux 
rose rather quickly to a high level accompanied by many dipping episodes (e.g.,~\citealt{shirey1999}). Compared to the
flux levels in previous years, this rate of the source even in this phase is now quite moderate
and also reduced by almost two orders of magnitude.

\subsection{Continuum Spectra and Fluxes
\label{cfluxes}}

In order to determine the source flux during the two exposures we fit the source spectra with
generic continuum models. Unlike in obs.~I and II, where we had severe 
spectral degradation due to photon pileup, we do not have to restrict ourselves to specific
spectral ranges or rely on higher orders in our new observations. While obs.~III
is practically pileup free, possible pileup at the maximum photon fluxes in obs.~IV is below 
2$\%$, which
we treat as a contribution to the systematic uncertainty. For the flux analysis we also 
exclude detected line features (see below) which although not significantly contributing to the 
overall source flux will affect the broad band continuum fit. The main motivation for this portion of the
analysis is to obtain X-ray fluxes which serve as 
a check on the outcome of the fits using the detailed photoionization model
below. We also investigate potential similarities of the continuum shape with previous 
observations.

\begin{figure*}
\includegraphics[angle=0,width=16cm]{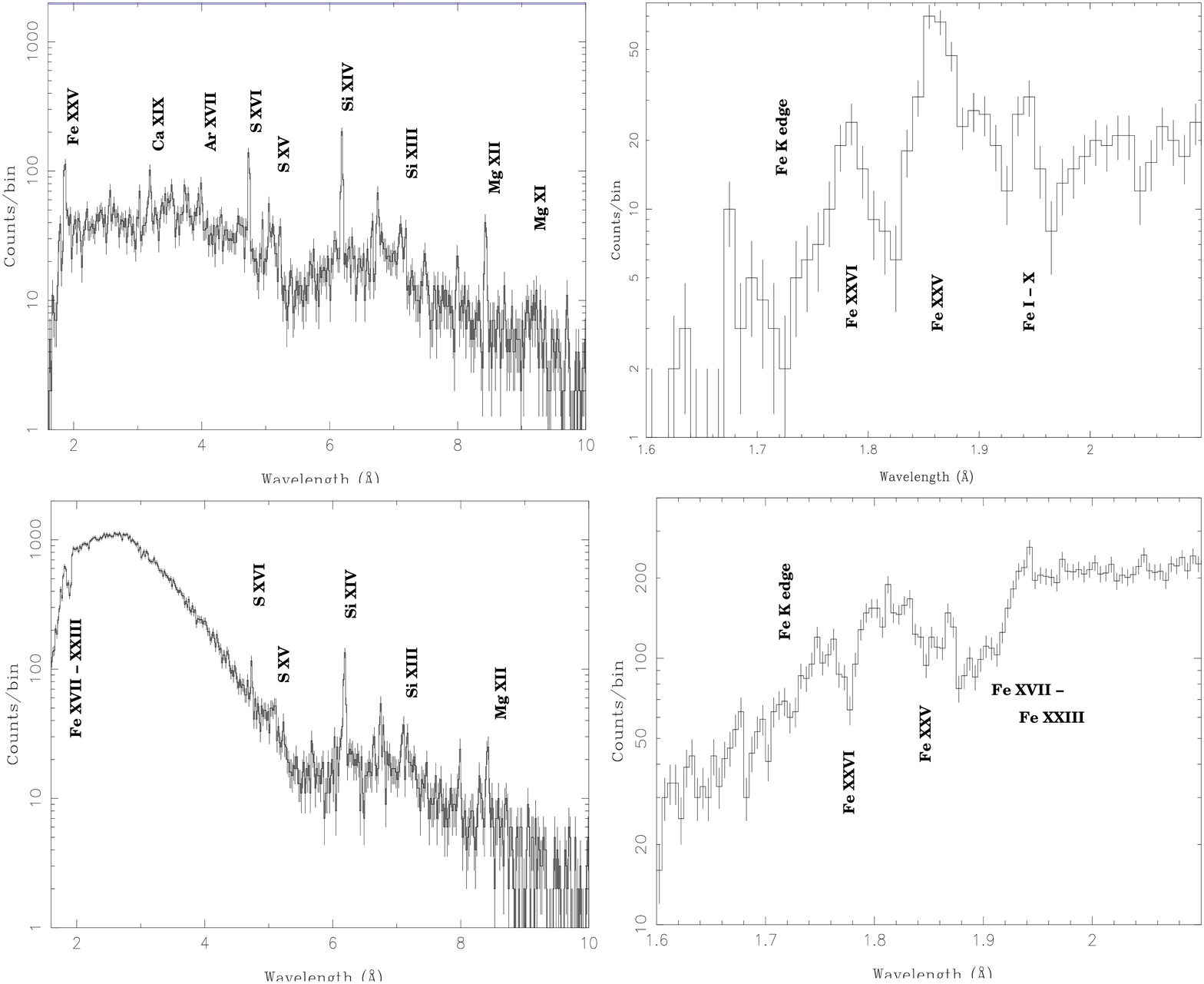}
\caption{The raw spectra of obs.~III (top) and IV (bottom) from all
HETG 1st order photons. In both cases the spectral bandpass is limited to 
wavelengths below 10~\AA~indicating high column densities.
\label{rawspectra}}
\end{figure*}
\smallskip
\smallskip

Figure~\ref{rawspectra} indicates that there is
no significant flux longward of 10~\AA. This is different from the previous observations, where
we have flux contributions up to 14~\AA~and use a model spectrum based on previous 
fits to \asca data ~\citep{brandt1996}
consisting of two blackbody components with both intrinsic partial-covering absorption
in addition to interstellar absorption. The raw spectra in obs.~III and IV indicate 
a high level of complexity involving
a combination of emission and absorption. We could not produce good fits for both observations
using the double partially covered blackbody model, but found
similar options. Table 2 summarizes the fit results.
Statistically only one simple power law plus absorption was required for  
observation~III. This was not the case in obs.~IV, which in order to include
the softer end of the spectrum (i.e., above 5.2~\AA) needed an additional component. 
\citet{galloway2007} successfully fitted a single blackbody with partial coverage absorption
to high flux spectra at higher orbital phases. The fits produced reduced $\chi^2$ below 1.5
only under certain conditions. In obs.~III the fit only allowed fdor interstellar absorption
leaving significant residuals near the Si and Fe K edges. In observation~IV the model fit only
if the wavelength range was restricted to 2 -- 8~\AA. 

The next step involved a two-component model consisting of two power laws, also
with additional absorption. One of the
components required significant additional absorption with a partial covering
fraction above the 90$\%$ level. Substitutions of one of the components with blackbody functions
left the reduced $\chi^2$ well above 1. Still, the two-component configuration is relatively similar
to what we found in observation~I and II (SBII) and we therefore applied it to
obs.~III as well. We found a good fit with rather similar parameters. Here the
substitution of the absorbed power law with a blackbody worked as well. The conclusion we
draw from this is that the new spectral shape is likely not much different from the one
in the high-flux states a few years ago. The models also suggest that the transition
between pre- and post-zero phases is predominantly a change in the hard absorbed spectral component.
Most important, however, is the fact that the continuum is consistent with a power law
shape in both, the hard and softer parts of the spectra, which then allows a relatively
straightforward treatment of photo-ionization and -absorption. The absorbed source flux (2 -- 10 keV)
in obs.~III was determined to (1.23$\pm0.02)\times10^{-10}$ \ergcm and in obs.~IV
was (5.72$\pm0.03)\times10^{-10}$ \ergcm. The interstellar
column density  is assumed to be 1.94$\times10^{22}$ cm$^{-2}$ as provided by many previous X-ray analyses
and which is close to the predicted column from H~I measurements in this line of sight~\citep{predehl1995}.
Both spectra in all models show that substantial additional absorption is required.

\begin{table*}
\begin{center}
{\sc TABLE 2: THE CONTINUUM FIT RESULTS} \\
\begin{tabular}{llcccccccc}
            & & & & & & & & & \\
\tableline
 Obsid &  Model & N$_{H}$ & N$^{pc}_{H}$ & f$_{pc}$ & $\Gamma_1$  & A$_1$ & $\Gamma_2$ or kT$_{bb}$ & A$_2$ & $\chi_{\nu}^2$ \\
       &        & [10$^{22}$ cm$^{-2}$] & [10$^{22}$ cm$^{-2}$]  &  &  & [cts s$^{-1}$ $\AA^{-1}$] & [], [keV] & [cts s$^{-1}$ \AA
$^{-1}$]& \\
\tableline
            & & & & & & & & &\\
 6148 & P      & 3.86$^{+0.30}_{-0.29}$ & --  & -- & 2.11$^{+0.16}_{-0.15}$  & 2.3$^{+0.6}_{-0.5}\times10^{-2}$ & -- & -- & 1.14 \\
      & PCBP & 2.86$^{+0.07}_{-0.08}$ & 4.46$^{+0.45}_{-0.64}$ & 0.95$^{+0.02}_{-0.03}$ & 1.40$^{+0.03}_{-0.16}$  & 6.8$^{+0.2}_{-1.8}\times10^{-3}$ & 0.43$^{+0.05}_{-0.03}$ & 6.1$^{+2.8}_{-0.6}\times10^{-3}$ & 0.90 \\
      & PCPP & 1.88$^{+0.06}_{-0.07}$ & 1.84$^{+0.08}_{-0.05}$ & 0.92$^{+0.04}_{-0.02}$ & 1.87$^{+0.03}_{-0.01}$  & 1.6$^{+0.6}_{-0.1}\times10^{-2}$   & 7.08$^{+2.07}_{-1.07}$ & 5.3$^{+6.0}_{-0.3}\times10^{-3}$ & 0.94 \\
 5478 & PCPP & 5.92$^{+0.05}_{-0.08}$ & 12.67$^{+0.09}_{-0.11}$ & 0.95$^{+0.01}_{-0.01}$ & 1.67$^{+0.00}_{-0.01}$ & 0.26$^{+0.00}_{-0.00}$       & 5.99$^{+0.26}_{-0.22}$ & 8.4$^{+0.6}_{-1.7}\times10^{-2}$ & 0.84 \\
            & & & & & & & & &\\
\tableline
\end{tabular}
\end{center}
\footnotesize
P = Power law; PCBP = partially covered blackbody + power law ; PCPP = partially covered power law + power law \hfill\break
\end{table*}

\section{Detailed Spectral Analysis}

Our models for the spectral features (emission and absorption lines,
and bound-free absorption) consist of both
phenomenological fits, using Gaussian lines, and fits to photoionization
models constructed using the XSTAR (v. 2.1ln2) code.\footnote{http://heasarc.nasa.gov/docs/software/xstar/}
XSTAR calculates both the ionization/excitation of a gas illuminated
by a strong source of X-rays and also the spectral features imprinted
by the gas, both in emission and absorption.  This is available as an
`analytic' model in XSPEC\footnote{http://heasarc.nasa.gov/docs/software/xspec/},
which can also be imported into ISIS.\footnote{http://space.mit.edu/CXC/ISIS/}  The
models assume a time-steady balance between processes affecting
ionization, recombination,
excitation, de-excitation, heating and cooling of the gas.
The key parameters determining the emission and absorption
properties are the shape of the incident spectrum and the ionization parameter,
defined as $\xi=4 \pi f_x / n_e$, where
X-ray flux $f_x$ is defined as  the energy flux incident on the gas integrated
from 1 -- 1000 Ry~\citep{tarter1969} and $n_e$ is the electron number density in the gas.
Additional parameters include the elemental abundances,
bulk velocity or redshift, and turbulent velocity, in addition to the
column density or the emission measure for absorption or emission models,
respectively.  Each model component is idealized as a uniform medium with a
given set of input parameters.  Emission and absorption components
can be combined, either singly or multiply, using standard
`model' commands in XSPEC or ISIS.

For all spectra we performed simultanous fits to the individual 1st order spectra of the MEG and
HEG. Only for plotting purposes are the spectra and model fits added together. 
During the analysis we binned the spectra into 0.005~\AA\ bins which matches the available MEG bins
and undersamples the HEG by a factor
of 2. 

\subsection{Emission Lines\label{emission}}

Table 3 lists all detected lines with positions and fluxes obtained
from Gaussian fits.  The values were obtained by fixing the continuum 
level to the fits from Section~\ref{cfluxes}. The line lists for the two observations 
differ substantially due to the fact that in obs.~IV we do not find any line emission
below 3.9~\AA\ except for traces of an Fe K fluorescence line. In obs.~III we detect
all major H- and He-like Lyman $\alpha$ lines plus a few weak H-like Lyman $\beta$ lines from some elements.
The fits to both spectra indicate no systematic or significant line shifts. Observation~III 
appears to be slightly inconsistent with this assessment. 
The laboratory wavelengths tabulated in column 2 are taken from tables in
the Atomdb web-guide. The difference 
in measured wavelength from the laboratory values for observation III distribute fairly well
around a centroid shift of 0.0045~\AA, which is well within the statistical and systematic uncertainties 
expected for this data set. In obs.~IV there are fewer lines but they distribute well around
the expected rest wavelengths.

The line shapes are resolved but require little broadening beyond the resolution of the HEG.
Most Doppler broadening line velocities are less than 800 km s$^{-1}$, except for Fe and some Mg lines.
The 90$\%$ confidence limits indicate that the lines themselves are just
barely resolved. There is no obvious correlation with wavelength scale except for the Fe region
where the resolution is already over 1500 km s$^{-1}$. The values for Mg stand out
a bit. Specifically the fact that we could not resolve the Mg XI triplet indicates some higher
velocity broadening. Since in both observations the lines are just barely resolved it is also
not obvious if the trend to lower velocities in obs.~IV is statistically relevant. Although
the lines tend to be brighter, the fits suffer from lower contrast due to a higher continuum.

Another peculiarity between the line emission of obs.~III and~IV is the fact that we do not detect
any emission lines with wavelengths shorter than Ar~XVII in observation~IV. Using the 1 $\sigma$ statistical
uncertainty of the underlysing continuum we limit the line fluxes for the undetected
line top below 1.2$\times10^{-6}$ ph s$^{-1}$ cm$^{-2}$. While we do not detect Ar~XVIII and Ca~XIX in
either absorption or emission, we detect
Ca~XX, Fe~XXV, and Fe~XXVI in absorption. Furthermore there is an unresolved line absorption complex
longward of Fe~XXV, likely due to lower transitions of Fe~XX to~XXII indicating the presence
of both highly ionized and moderately ionized absorbers in the line of sight.
We refer to these respectively as `hot' and `warm' components,
although these designations do not reflect the fact that the dominant source
of ionization is likely the continuum photons from the compact object
rather than thermal electrons. The line positions of  the hot
absorption lines do not deviate significantly from the laboratory position. This is different
to obs.~I and~II where they appeared blue-shifted, thereby producing P Cygni lines. A more detailed analysis
of the absorbers is given below.

\begin{table*}
\begin{center}
{\sc TABLE~3: X-RAY LINE PROPERTIES} \\
\begin{tabular}{lccccccc}
 & & & & & & \\
\hline
\hline
 & & & & & & \\
     & & \multicolumn{3}{c}{Observation~III} & \multicolumn{3}{c}{ Observation~IV} \\
 & & & & & & \\
\hline
 ion & $\lambda_o$ & $\lambda_{\rm III}$  & Flux$_{\rm III}$ & $v_{\rm III}$ &$\lambda_{\rm IV}$  & Flux$_{\rm IV}$ & $v_{\rm IV}$ \\
     & \AA  & \AA & 10$^{-4}$ ph s$^{-1}$~cm$^{-2}$ & km s$^{-1}$ & \AA &  10$^{-4}$ ph s$^{-1}$~cm$^{-2}$ & km s$^{-1}$ \\
\hline
Fe~XXVI           &   1.780 &   1.783$\pm$0.002 &  0.97$\pm$0.20 & 1680$\pm$1490 & 1.776$\pm$0.001 & $-2.45\pm$0.08 & 740$\pm$880 \\
Fe~XXVr           &   1.850 &   1.853$\pm$0.001 &  1.58$\pm$0.17 &  570$\pm$1130 & 1.848$\pm$0.002 & $-1.78\pm$0.34 &1090$\pm$850 \\
Fe~XXVi           &   1.859 &   1.860$\pm$0.002 &  0.80$\pm$0.15 & 1520$\pm$1200 & -- & -- & -- \\
Fe~XXVf           &   1.868 &   1.870$\pm$0.001 &  1.07$\pm$0.15 &  570$\pm$1120 & -- & -- & -- \\
Ca~XX~L$\alpha$   &   3.021 &   3.023$\pm$0.002 &  0.26$\pm$0.06 &  210$\pm$450 & 3.019$\pm$0.001 & $-0.83\pm$0.19 & 430$\pm$520 \\
Ca~XIXr           &   3.177 &   3.163$\pm$0.012 &  $<$0.07        &  $<$650 & -- & -- & -- \\
Ca~XIXi           &   3.189 &   3.184$\pm$0.001 &  0.51$\pm$0.15 &  660$\pm$950 & -- & -- & -- \\
Ca~XIXf           &   3.211 &   3.205$\pm$0.003 &  0.25$\pm$0.10 &  590$\pm$650 & -- & -- & -- \\
Ar~XVIII~L$\alpha$&   3.734 &   3.738$\pm$0.005 &  0.40$\pm$0.13 &  590$\pm$990 & -- & -- & -- \\
S~XVI~L$\beta$    &   3.780 &   3.787$\pm$0.006 &  0.24$\pm$0.12 &  550$\pm$470 & -- & -- & -- \\
Ar~XVIIr          &   3.949 &   3.939$\pm$0.004 &  $<$0.06        &  $<$610      & 3.932$\pm$0.005 & $<$0.01        & $<$660 \\
Ar~XVIIi          &   3.966 &   3.956$\pm$0.004 &  0.24$\pm$0.13 &  360$\pm$530 & 3.958$\pm$0.007 & 0.19$\pm$0.18 & 130$\pm$110 \\
Ar~XVIIf          &   3.994 &   3.995$\pm$0.003 &  0.45$\pm$0.15 &  600$\pm$520 & 3.998$\pm$0.005 & 0.34$\pm$0.19 & 600$\pm$490 \\
S~XV~L$\beta$     &   4.310 &   4.310$\pm$0.006 &  0.28$\pm$0.15 &  610$\pm$220 & -- & -- & -- \\
S~XVI~L$\alpha$   &   4.730 &   4.737$\pm$0.001 &  2.12$\pm$0.24 &  540$\pm$320 & 4.734$\pm$0.003 & 1.82$\pm$0.40 & 420$\pm$390 \\
S~XVr             &   5.039 &   5.046$\pm$0.006 &  0.34$\pm$0.24 &  250$\pm$170 & 5.042$\pm$0.008 & 0.40$\pm$0.23 & 210$\pm$200 \\
S~XVi             &   5.067 &   5.056$\pm$0.007 &  0.40$\pm$0.30 &  780$\pm$440 & 5.070$\pm$0.015 & 0.31$\pm$0.27 & 210$\pm$200 \\
S~XVf             &   5.102 &   5.106$\pm$0.005 &  0.44$\pm$0.15 &  720$\pm$910 & 5.103$\pm$0.006 & 0.60$\pm$0.26 & 270$\pm$260 \\
Si~XIV~L$\beta$   &   5.217 &   5.222$\pm$0.005 &  0.55$\pm$0.25 &  390$\pm$510 & 5.225$\pm$0.007 & 0.53$\pm$0.30 & 300$\pm$330 \\
Si~XIV~L$\alpha$  &   6.183 &   6.192$\pm$0.001 &  1.54$\pm$0.18 &  550$\pm$110 & 6.185$\pm$0.002 & 3.43$\pm$0.47 & 490$\pm$230 \\
Si~XIIIr          &   6.650 &   6.662$\pm$0.005 &  0.24$\pm$0.11 &  280$\pm$310 & 6.643$\pm$0.007 & 0.69$\pm$0.33 & 390$\pm$480 \\
Si~XIIIi          &   6.669 &   6.690           &  $<$0.11        &  $<$310      & 6.705$\pm$0.010 & 0.21$\pm$0.25 & 160$\pm$580 \\
Si~XIIIf          &   6.740 &   6.746$\pm$0.005 &  0.30$\pm$0.09 &  800$\pm$410 & 6.741$\pm$0.003 & 1.12$\pm$0.34 & 290$\pm$290 \\
Mg~XII~L$\beta$   &   7.106 &   7.104$\pm$0.012 &  0.36$\pm$0.13 & 1450$\pm$1100& 7.109$\pm$0.007 & 1.07$\pm$0.40 & 690$\pm$530 \\
Mg~XII~L$\alpha$  &   8.422 &   8.432$\pm$0.005 &  1.40$\pm$0.33 &  670$\pm$200 & 8.422$\pm$0.007 & 4.65$\pm$1.43 & 690$\pm$370 \\
Mg~XI(r,i,f)      &   9.230 &   9.219$\pm$0.084 &  2.31$\pm$1.01 & 3940$\pm$680 & -- & --  & -- \\
\hline
\end{tabular}
\end{center}
\end{table*}

The Fe K region shows some peculiarities as well.  In observation III
the Fe XXV is clearly the dominant feature.  It does show some excess flux
longward of the forbidden line position, however with low statistical significance. 
Under the assumption that we are observing
a photoionized plasma, then these could be trace contributions from inner
shell transitions of Fe XIX -- XXII ions~\citep{kallman2004}.  If so, this would suggest
the presence of a distribution of ionization parameters,
extending to values $\sim$10 times less than the dominant
component.

The main challenge for the photoionization model comes from the He-like lines,
here Fe~XXV, Ca~XIX, Ar~XVII, S~XV and Si~XIII and Mg~XI, observed in obs~III.
In interpreting these lines we noted their dependence
on gas density and excitation mechanism ~\citep{bautista2000, porquet2000}:
each line complex consists of three components,
denoted r, i, f, ordered in increasing wavelength.  At densities
below the critical density we expect R=f/i$\geq$1, and conversely.
Critical densities depend on the nuclear charge, $Z$, and range from $\sim 10^6$
cm$^{-3}$ for C to $\sim 10^{17}$ cm$^{-3}$ for Fe.
Gas that is photoionized is expected to emit these lines primarily by
recombination cascade, and will have G=(f+i)/r$\geq$1.  Gas which is
excited by electron collisions, or in which resonantly scattered emission
is important, will have G$\leq$1.
The He-like complexes in obs.~III all provide us with constraints
on the excitation mechanism and gas density, although taken together they
provide conflicting indications of these quantities.
The Fe XXV feature peaks near 1.86~\AA, which is the rest energy of the
intercombination line, but it extends beyond 1.87~\AA, the energy of the
forbidden line, and there is little emission at 1.85~\AA, the energy of the
resonance line.  The Ca XIX line behaves similarly.  We only measure an
upper limit to the resonance line, but find strong intercombination and
weak forbidden lines.  However, there is excess emission at wavelengths
greater than the laboratory wavelength of the forbidden line, which cannot
be adequately fitted by the XSTAR model.  Ar XVII is an intermediate
case in which the fit accepts both a resonance line and a forbidden line.  S XV is an
intermediate case as well, though its line components appear less significant.
Si XIII is a clearer case, and we have data from both observations.  Figure~\ref{sitriplets}
shows the spectrum in both cases, in which the resonance and forbidden lines
are clearly detected and the intercombination lines are upper limits.
This suggests a density below the critical density for the
2s$^3$P -- 2s$^3$S transition, which is 5 $\times 10^{14}$ cm$^{-3}$ for
Si.  The He-like line ratios from the heavier elements suggest that if there
is a single plasma density it is likely near that limit.
Although we cannot resolve Mg~XI in obs.~III, the line clearly peaks at the
location of the intercombination line indicating an R ratio well below 1 and
thus densities significantly above $10^{13}$ cm$^{-3}$ (see also the discussion in
Sect. 4.1).

\includegraphics[angle=0,width=8.5cm]{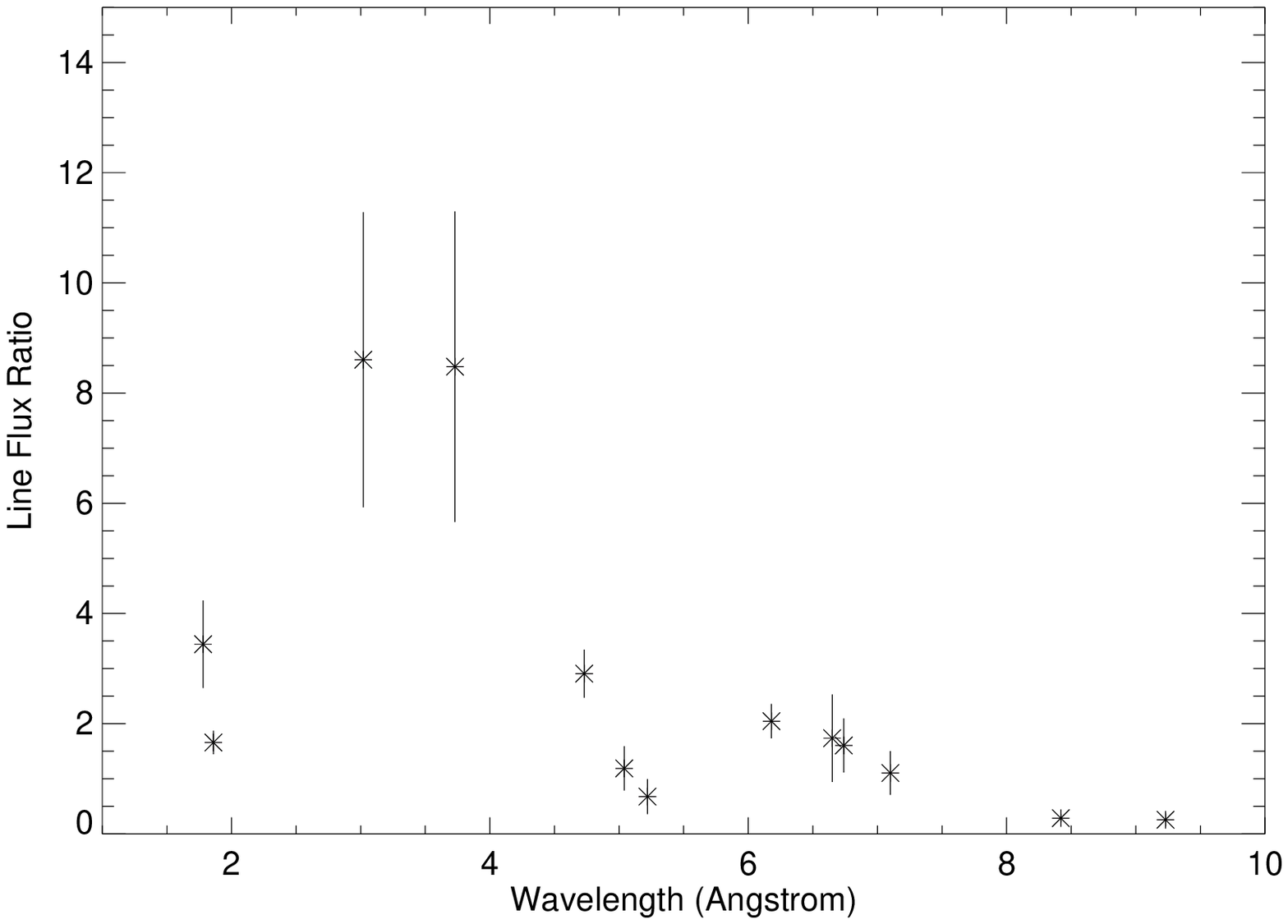}
\figcaption{Flux ratios $f_{II}/f_{III}$ for lines measured in obs.~III and II (values taken
from SBII).
\label{lratios}}
\smallskip
\smallskip

The fluxes of the emission lines are shown in Table 3. The lines in obs. IV
appear bright with fluxes between 2 $\times 10^{-5}$ and
3.4 $\times 10^{-4}$ ph cm$^{-2}$ s$^{-1}$, in obs. III
ranging from 1.5 $\times 10^{-5}$ ph cm$^{-2}$ s$^{-1}$ to
2.1 $\times 10^{-5}$ ph cm$^{-2}$ s$^{-1}$. 
The line fluxes between the two observations are not substantially different from each other, most 
pre-zero phase fluxes seem slightly lower. This is different from what we observed during
the high flux states (SBII) where the line fluxes differed substantially. The flux
difference between pre- and post-zero phase then also was an order of magnitude larger than
what we observe now. Another issue arises when we compare the pre-zero phase lines 
between obs.~II and~III (see Figure~\ref{lratios}). The reason we use 
observation~II is because here we expect the least perturbing effects of pileup on the 
lines. The source flux in obs.~II was determined to be $4.8\times10^{-9}$~\ergcm which
leads to a continuum flux ratio of about 150. In case that we observe a similar photoionization
regime, we would expect that the line ratios are very similar to this
flux ratio. The ratios in Figure~\ref{lratios} appear much smaller than this expectation. 
Some of this effect is certainly caused by pileup in obs.~II which likely causes the
slope from large to small wavelengths. Even with some of the line fluxes reduced
by pileup effects,  
the line ratios fall still well short of the expectation from the continuum ratios, which 
besides the effects of the now reduced source luminosity we attribute to
additional absorber activity as well as a more dramatic change in ionization balance. 

\smallskip
\smallskip
\includegraphics[angle=0,width=8.5cm]{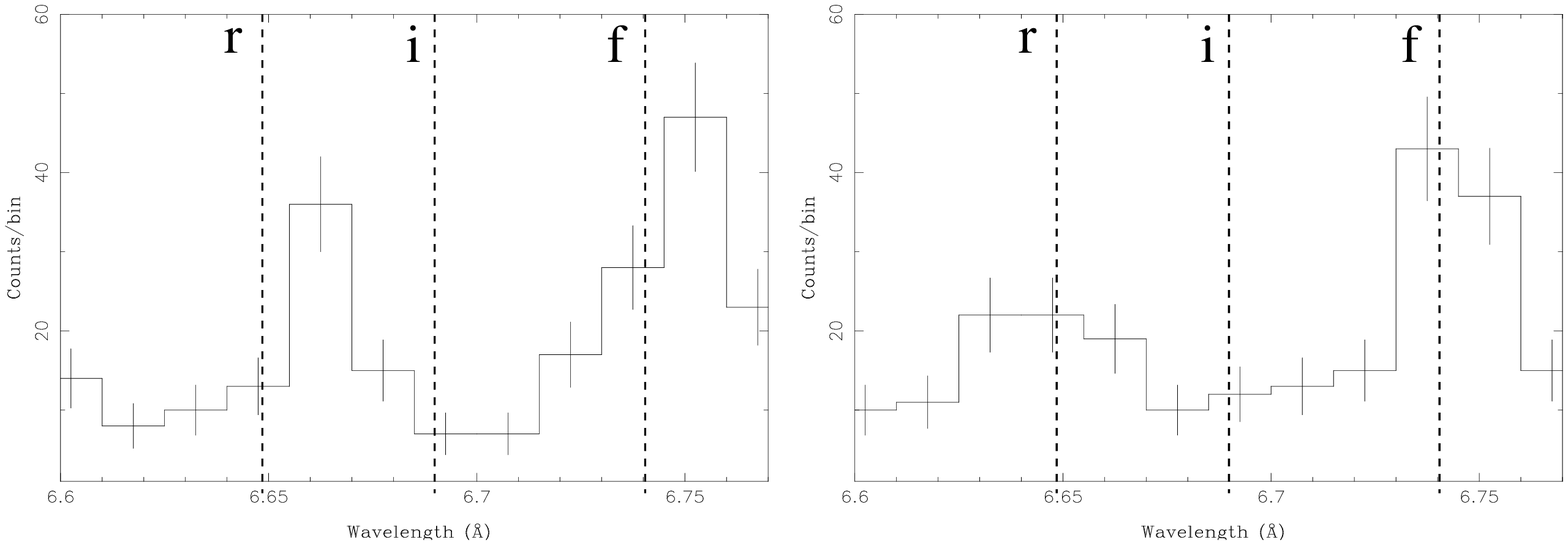}
\figcaption{The Si XIII triplet from Obs. III (left) and IV (right). In both triplets the forbidden
line is the strongestxi and  the intercombination line is weak. The slight redshift for obs.~III is within the
systematic uncertainty of the wavelength scale.
\label{sitriplets}}
\smallskip
\smallskip

\subsection{Photoionization Fits with XSTAR\label{photo}}

The spectral analysis based on Gaussian fits indicates complex ionization structures.
First we investigate photoionization in observation~III. For the XSTAR fits we use the PHOTEMIS analytic
model in XSPEC, which is available from the XSTAR 
website.~\footnote{http://heasarc.gsfc.nasa.gov/docs/software/xstar/xstar.html} 
This model  generates a line spectrum characterized by element abundances, velocity turbulence,
ionization parameter and normalization. Density is not a free parameter in the fits, 
but it can be varied by importing XSTAR model runs calculated at various density values. Temperature is 
also not a free parameter in these fits, since it is calculated self-consistently by XSTAR
under the assumption that heating due to photoionization is balanced by the radiative cooling of the gas.
The determination of the ionization parameter is of the most interest in this analysis.

\begin{table*}
\begin{center}
{\sc TABLE 4: THE FIT RESULTS USING THE PHOTOIONIZATION (PH) AND WARMABSORBER (WA) MODEL IN XSTAR} \\
\begin{tabular}{lcccccccc}
            & & & & & & & & \\
\tableline
 Obsid &  Model & log N$^{cold}_{H}$ & log N$^{warm}_{H}$ & $\xi$ & $\Gamma$  & z & f$_x^a$ & L$_x^b$ \\
       &        & [cm${-2}$] & [cm${-2}$]  & [ergs cm s$^{-1}$] &  &  & & [$10^{36}$ \ergsec]\\
\tableline
            & & & & & & & & \\
 6148 & PH & 22.9$^{+0.0}_{-0.0}$ & --  & 3.0 & 1.58$^{+0.00}_{-0.00}$  & 0.001 & 1.219 & 0.79 \\
 5478 & WA (quies.) & 22.3$^{+0.1}_{-0.1}$ & 23.9$^{+0.1}_{-0.0}$  & 1.6$^{+0.4}_{-0.2}$ & 0.38$^{+0.29}_{-0.19}$ & -0.0077$^{+0.0028}_{-0.0047}$ & 1.51 & 1.4 \\
 5478 & WA (flar.)& 23.4$^{+2.4}_{-0.7}$ & 22.8$^{+0.1}_{-0.1}$  & 2.7$^{+0.1}_{-0.2}$ & 2.62$^{+0.12}_{-0.12}$ & -0.0019$^{+0.0019}_{-0.0021}$ & 8.31 & 9.9 \\
\tableline
\end{tabular}
\end{center}
\footnotesize
a) $\times 10^{-10}$ \ergcm between 2 -- 10 keV \hfill\break
b) for a distance of 6 kpc \hfill
\end{table*}
 
The model spectrum produces all the lines listed in Table 3 and reproduced the listed fluxes
within the quoted errors for over 75$\%$ of the lines. 
The lines are slightly broadened (see above) which was reflected in a turbulence velocity parameter
of 300 km s$^{-1}$.  
The temperature of the photoionized plasma is largely set by the detection
of radiative recombination continua. Since there are no significant detections of these, the temperature
of the plasma must be greater than 10$^6$ K which blends these features into the continuum;
the temperatures of the XSTAR models at the best-fit
ionization parameter are consistent with this constraint.
The density of the plasma is determined by the He-like triplets. As indicated in Sect.~\ref{emission},
the triplets provide conflicting trends. The fit allows a lower density solution as suggested 
predominately by the Si XIII triplet with $\sim 10^{14}$ cm$^{-3}$, which is only in conflict with the
density constraints from
Ca XIX and Fe XXV, which seem to suggest $> 10^{15}$ cm$^{-3}$. An ionization parameter ($\xi = L_x/(n r^2)$) of 
log$\xi$=3.0$\pm$0.1 gives the
best result and provides a cold column of 7.9$\pm0.1 \times 10^{22}$ cm$^{-2}$, a power law index
of $\Gamma = 1.58\pm0.1$, and a total X-ray flux of 1.22$\times10^{-10}$ \ergcm.  
The fitted flux and power law index are consistent with 
our continuum analysis.
XSTAR also allows some constraint on abundance and although some slightly higher abundances
are seen for Si and Ca, the fit clearly required significantly enhanced Fe.   

\subsection{Line Emissivities and Predictions\label{adcmodels}}

It is straightforward to explore the implications of our measurements
of emission line fluxes for models for the accretion flow in
Cir X-1.  For example, from the Gaussian fit to the lines we find a net flux
of approximately 7.9 $\times 10^{12}$ erg cm$^{-2}$ s$^{-1}$ in the
lines making up the Fe XXV K complex.  At a distance of 6 kpc this
corresponds to a line luminosity of 3.2 $\times 10^{34}$ erg s$^{-1}$.
If this were emitted by recombination in a region with uniform conditions, the luminosity would be:

\begin{equation}
L_{line}= n^2 V \alpha_{2p} \varepsilon_{line} x_{Fe XXVI} y_{fe}
\end{equation}

\noindent where $n^2 V$ is the emission measure, $\alpha_{2p}$ is the effective recombination rate into
the $2p$ levels of Fe XXV, $\varepsilon_{line}$ is the line energy  $x_{Fe XXVI}$ is the ion fraction
of Fe XXVI and  $y_{fe}$ is the iron elemental abundance.  From XSTAR models
we have $\alpha_{2p} x_{fe XXVI} = 1.95 \times 10^{-12}$ cm$^3$ s$^{-1}$.
Combined with the measured line luminosity, we find that the emission measure at log($\xi$)=3 must therefore
be $n^2 V = 4.9 \times 10^{58}$ cm$^{-3}$.

This quantity can be compared with what is expected from an X-ray heated accretion disk coronae (ADC) in low mass
X-ray binaries.  Simple estimates for the emission measure distribution from
an ADC are complicated, owing to the fact that the coronal temperature, density and ionization are coupled
by the constraints of hydrostatic equilibrium, thermal balance, and the
radiative transfer of photons into the corona.  Calculations for the ADC structure including these effects have been
published by ~\citet{london1982, begelman1983, kallman1989, ostriker1991, ko1994, rozanska2002, mario2002},
and many others.  A simple estimate for the distribution of emission measure
with ionization parameter

\begin{equation}
\frac{d EM}{d\xi}=\int{n^2 2\pi R dR \frac{dz}{d\xi}}
\end{equation}

\noindent can be obtained using an isothermal approximation to the
disk vertical structure:  $n(z)=n_0 exp(-(z/z_s)^2)$,
where the scale height is $z_s=\sqrt(kTR^3/(GMm_H))$.
Then:

\begin{equation}
\frac{{d EM}}{d\xi}\simeq \xi^3 \left(\frac{kT}{GMm_HR_i^3}\right)^{1/2}L^2
=2.5 \times 10^{65} \xi^{-3} L_{38}^2 R_{i7}^{-3/2}T_4^{1/2}
\end{equation}

\noindent where $L_{38}$ is the source luminosity in units of $10^{38}$ erg/s, $R_{i7}$ is the disk inner radius in units of
$10^7$ cm and $T_4$ is the corona temperature in units of $10^4$ K.

\includegraphics[angle=-90,width=8.5cm]{f6.eps}
\figcaption{Calculation of emission measure in dependence of ionization parameter for a static corona for two different
luminosities, $L=10^{38}$ \ergsec (red) and $L=10^{37}$ \ergsec (green).
\label{corona}}
\smallskip
\smallskip

This result is quite accurate when compared with the results of numerical solutions for
a single-stream treatment of the transfer of the X-rays, and for
a disk around a 1 $M_{\odot}$ compact object extending from $10^7$ to $10^{11}$ cm.  Figure~\ref{corona} shows
such a calculation, for $L_{38}$=1 (red curve) and $L_{38}$=0.1 (green curve). This shows that at log($\xi$)=3,
an emission measure that we infer from the observations can only be produced by an ADC if the source
luminosity is well above $10^{37}$ erg/s. Although this estimate of 
emission measure is idealized, and assumes a static corona above a 
thin disk, it is clear that the continuum luminosity needed to produce the 
line emission we observed exceeds the luminosity we 
infer from continuum fits by a factor almost two orders of magnitude.  One possible explanation 
for this is that the true luminosity intrinsic to the source is much 
greater than what we observe, and that the apparent low state is a 
consequence of obscuration or collimation of the radiation into a 
direction away from our line of sight.

\subsection{The warm absorber in observation~IV}

Figure~\ref{hardlines} clearly indicates that the hard spectrum in obs.~IV has not only changed
in its continuum properties, but also exhibits complex absorber activity. 
Encouraged by the similarity of the line spectra above about 4~\AA\ in obs.~III and IV we use
the results from Sect.~\ref{photo} in the modeling of the post-zero phase spectra.
We adopt two XSTAR components to model the
spectrum in obs.~IV, the warm absorber model (WARMABS) plus the photoemission model (PHOTEMIS)
determined from observation~III. The data quality is not sufficient to constrain all the parameters of both components
unambiguously, so in order to reduce the number of free parameters we fix the photoemission
component to the one fitted in observation~III. This  
choice also makes sense under the presumption that dips are generated by blocking the central
X-ray source and thus exposing the 
accretion disk coronal spectrum as observed in, for example, dipping sources
like EXO 0748-676~\citep {mario2003}.
The foremost photoionization parameters we freeze from the above analysis are the ionization parameter
for the photoemission, the
normalization and the turbulence velocity parameter. 
In a second step we recognize that the obs.~IV shows significantly  more variability
than obs.~III. We then separated the quiescent segments in the light curve of 
obs.~IV (see Figure~\ref{lightcurves}) and the flaring segments. 

\includegraphics[angle=-90,width=8.5cm]{f5.eps}
\figcaption{The total spectrum below about 5~\AA~from obs.~IV showing photoemission of the soft side
but complex photoabsorption on the hard side.
\label{hardlines}}
\smallskip
\smallskip

In the following we denominate the two levels accordingly as ``quiescent'' and ``flaring''
states. 
Table 4 shows the results of the XSTAR fits to the quiescent and flaring
state. The model fits well to both states and produces significantly different results.
During the quiescent state the warm absorber model results in a low
ionization parameter of  $log~\xi = 1.6 \pm 0.3$. The absorption feature at Fe K then corresponds to 
absorption from Fe XVI -- XVIII. The cold column seems reduced with respect to the 
one observed in obs.~III, but a warm column of 7.9$\times 10^{23}$ cm$^{-2}$ appears.
In the flaring state the cold column is reduced as well with a lower warm column
of 6.3$\times 10^{22}$ cm$^{-2}$ but a high ionization parameter of $log~\xi = 2.7 \pm 0.2$.
The absorber near Fe K has its strongest contributions now from Fe~XXI-XXIII. 
In order to obtain the best fit a blueshift of
approximately 600 km s$^{-1}$ is adopted.  However, this value
yields a fit which is only marginally improved ($\Delta\chi^2\simeq$3)
compared with no blueshift.  Similarly, tests of the fits to the
obs. IV data with and without the emission component show  marginal
improvement ($\Delta\chi^2 =$8 for the flaring state and $\Delta\chi^2 =$6 
for the quiescent state). Table 4 allows lists the total luminosities for the two WA cases.

Figure~\ref{depth} shows the dependence of the depths of observed absorption lines
on ionization parameter for the flaring state. The fiducial optical depth is
chosen arbitrarily, i.e. the column density of the absorber. Plotted is the log
of the ratio of the optical depth produced by an XSTAR model with
total column density 10$^{22}$ cm$^{-2}$ and solar abundances~\citep{grevesse1999} 
to the observed depths. Most curves intersect near log $\xi \sim 2.7$.

These were determined using a simple Gaussian fit to the lines, and the plotted error bars
reflect the statistical error associated with these fits. 
This shows the range of $\xi$ which is approximately consistent with the observations,
and also illustrates the nature of the constraints:  the Fe XXVI line 
constrains ionization parameter on the low side, and the Si and S lines
constrain the ionization parameter on the high side. 

\includegraphics[angle=-90,width=8.5cm]{f7.eps}
\figcaption{The dependence of the depths of observed absorption lines
on ionization parameter for the XSTAR fits in the flaring state.
\label{depth}}
\smallskip
\smallskip

\section{Discussion}

The at times ultraluminous X-ray source \circinus has been observed by nearly all
X-ray missions over
the last 30 years, and it seems that the source continues to provide a plethora of new
features in its X-ray emission. The light curves, as expected from the monitoring with 
the ASM, show much less activity before but also after zero orbital phase. This 
confirms that the source has moved into a novel physical state. In fact, we suspect
that \circinus was still in transition because since January 2007 until the submission
of this paper the X-ray flux dropped under the detection limit of the ASM. Whether the source
is in the process of turning off or reaching a defined minimum is unknown. We suspect
the latter if we follow the 30 yr light curve compiled by~\citet{parkinson2003}, though
this light curve does not provide fully continuous coverage during its previous low-flux periods
that lasted between 1971 and 1986. The analysis 
in this paper aims to identify the physical state of the X-ray emitting plasma and
draw conclusions about the source's condition while it transits towards lower fluxes.
We also put these findings into context 
with the sources' previous high-flux behavior.  

\subsection{Some General Remarks}

There are still many outstanding questions concerning the detailed nature of the system. 
Besides its identification
as an accreting neutron star~\citep{tennant1986}, there is circumstantial evidence that \cirx1
could be a Z-type LMXB~\citep{schulz1989, hasinger1989} based on its spectral variation pattern
in the color-color diagram (e.g.,~\citealt{shirey1999, ding2006}). However, given recent findings, it has to be 
a quite peculiar one for several reasons. Its long-term light curve fits more the description
of an X-ray transient with excursions from 1 -- 2 Crab to less than a mCrab ~\citep{parkinson2003}.
Our observations, as well as recent \swift and \xte observations~\citep{jonker2007a}, show that Cir X-1
has gone from an all-time high during the 1990s into an extended low-flux state where the source dropped
to a flux of 3$\times10^{-11}$ \ergsec corresponding to a luminosity that is below 10$^{35}$ \ergsec.
LMXBs that low in luminosity are rare and  the detection of an LMXB close to the Galactic Center
with a luminosity of only 4$\times10^{34}$ \ergsec but high-luminosity outbursts~\citep{muno2005} 
may indicate the existence of a peculiar class of LMXB transients. Conventional Z-source do not
behave that way, and even atoll sources are brighter in their hard states. A most compelling 
indication that \cirx1 is indeed a neutron star is a  
recent detection of twin kHz quasiperiodic oscillations (QPOs) during the high-flux state
~\citep{boutloukos2006}, which seem in good
agreement with Z-sources, but not with black-hole binaries. 
Furthermore, the discussion about the companion has received attention again more recently as~\citet{jonker2007b} 
produced some evidence that the companion star of \cirx1 is of intermediate mass and of types
B5 to A0. This is unusual for a Z-source, as the ones known all have low-mass companions. 

~\citet{jonker2007b} constrain the companion mass to $< 10$~\Msun and the angle of inclination
to be $> 14^{\circ}$. The angle of inclination has been under dispute since evidence for
ultrarelativistic radio jets was presented~\citep{fender2004a}. An angle of $<5^{\circ}$ became
a natural choice under the assumption that the energy injection happened during the last
zero-phase outburst and that there is no bending. However, there is no evidence for this 
correlation. Neither recent claims of soft excesses~\citep{iaria2005} nor unabsorbed soft components
~\citep{ding2006} can be substantiated in our high-resolution spectra during the low-flux state
here and in the high-flux state years ago (see also SBII, \citealt{galloway2007}), and any claims
referring to the angle of inclination likely do not apply. In
\citet{ding2006} it was suggested that this excess could come from the accretion hot spot and
it large emission volume would then rule out a large inclination. However the spectral analysis
presented seems clearly tainted by an unrealistically high Fe K line flux. Besides the fact that
the continuum above the Fe K edge cannot physically produce such a line strength, we resolved 
Fe K lines in the high state and found them to be two orders of magnitude lower in strength (SBII).
This leaves us with the unfortunate reality that we are still quite uncertain about the angle of 
inclination.

If the result that the companion is of intermediate mass holds up, our analysis can very much
rule out a possible later B-type stellar wind as the region of origin for the X-ray lines.
Later type B-stars with surface temperatures above 10$^4$ K can still maintain substantial winds.
The size of the system is about 5$\times10^{12}$ cm~\citep{tauris1999} which for the measured ionization parameter
results in densities of the order of 4$\times10^6$ cm$^{-3}$, which even for a late B-types seems
quite low. The He-triplets suggest much higher densities than that which places the
ioniziation region much closer to the X-ray source. We also measure significantly enhanced Fe in the
photoemission, which is quite hard to argue for in an early type stellar wind.

\includegraphics[angle=-90,width=8.5cm]{f8.eps}
\figcaption{The range of R and G ratios for the 
He-like line emission in Obs.~III.
\label{density}}
\smallskip
\smallskip

Before we discuss emission regions in more detail, we have to address our difficulties
determining a plasma density from the He-like triplets. Though we are somewhat plagued by bad statistics
and some uncertain systematic effects, we still expected a more consistent picture. To date our ability to 
diagnose the high-density regime has always by limited by the unavailability of data and these observations
are a rare exception. The Mg~IX triplet indicates an R-ratio $<<$ 1 which puts a lower limit on the density of 
10$^{13}$ cm$^{-3}$. Figure~\ref{density} shows a cartoon relating the R-ratios with the corresponding
G-ratios. High-density regimes are a consistent solution for all triplets but Ar~XVII, which is however
statistically weak. Ca XIX, on the other hand, though it would be consistent with high density, does
not fit into the recombining regime and appears more collisional. Fe XXV provides the tightest solution
for higher density, but is also close to a collisional solution. Though all in 
all a density of $\sim5\times10^{14}$ cm$^{-3}$ in a recombining
gas seems most plausible, we want to caution that there are remaining inconsistencies, which, given the turbulent nature
of the system, could indicate time-dependent aspects and a situation where our equilibrium assumption
may not be entirely valid. 

\subsection{The Photoionized Disk Corona\label{photocorona}}

Quite evidently, the P Cygni lines found
during the early \chandra observing cycles (BSI, SBII) are not present
in the new spectra. So far these line profiles still stand unique for this class of objects.
The last time we observed these P Cygni lines was during a short observation
of \cirx1 at phase 0.25 during the year 2001~\citep{galloway2007}. The interpretation at the
time was that the profiles are a result of a preferentially equatorial wind driven
by the central source. In observation~III, the overall source flux was reduced
by over two orders of magnitude with respect to the 2000 observations and one of the questions
to answer is if this is a true reduction in source luminosity or an effect of
increased circumstellar/circumdisk absorber activity. Our direct comparison
of the ratio of the line fluxes with the ratio of the
continuum fluxes suggests that absorption is playing an important role, even though that comparison
is tainted by the pileup problems during the high-state observations.
On the other hand, photoionized plasmas are complex and we have to look at the change in ionization balance
between the high and now low flux phases represented by the ionization parameters.
Although we were not able to directly measure $\xi$ in obs.~I and~II we could infer
parameters well in excess of $10^4$ using calculations from \citet{kallman2001}. Our fits
to obs.~III resulted in an ionization parameter of $10^3$, which is an order of magnitude lower.
Thus significant dimming of the X-ray flux thereby changing the ionization balance provides
a solution that would not require significant additional absorption. Thus the comparison of line
and continuum fluxes and the comparison of ionization balances have quite different
implications. In that sense
the observed X-ray emission may not provide a full account of the true energy balance near the
central source and we have to acknowledge the fact that
there has been no significant long-term
increase in the observed cold column between the high and the low flux states for the years 2000 and
2005, respectively.  

At this point it seems necessary to invoke some model assumptions about accretion dynamics.
In order to test the assumption that the photoemission region could be the accretion disk we
compare results from our XSTAR fits of obs.~III to some simple model predictions. As a basis for these
tests we use some of the models described by ~\citet{mario2002}, who modeled an accretion-disk 
atmosphere and corona photoionized by a central X-ray continuum source and also discuss feedback mechanisms
between disk structure and the central illumination. In this model we assume an optically thick
standard disk \citep{shakura1973} with a heated atmosphere, which is separated from the disk by a
$\tau = 1$ surface, sometimes called photosphere. The atmosphere is strongly heated resulting in
evaporation into a hot corona and eventually leads to outflows. The plasma above the photoshere is
heated up to 10$^7$ K and cools through atomic line emission. The most important parameter to 
investigate the state of the plasma is the ionization parameter $\xi$. In order to determine
the source luminosity $L_x$ we assume a source distance of 6 kpc. The distance is not well known but it is 
widely agreed that the source is not closer than this. The claim by \citet{iaria2005} of 4.1 kpc has been
recently dismissed by \citet{jonker2007b} who actually favor a distance of $\sim$8 kpc. 
For the density $n$ we have some 
indication from the He-like triplets, but plausibility arguments are still needed.

In the inner disk under the idealistic assumption that viscous heating
and radiative heating from an illuminating source is locally expressed by a blackbody, 
radiative heating should dominate over viscous heating at radii larger than 2.3$\times10^8$ cm~\citep{vrtilek1990}.
With a luminosity of 8$\times10^{35}$ \ergsec we expect an upper limit to the  density
of $\sim 10^{16}$ cm$^{-3}$. The strength of the forbidden line in the Si XIII argues against such a
high density and if we employ our projection of $\sim5\times10^{14}$ cm$^{-3}$, we place the emission region
at a radius of $\sim 10^9$ cm. 
During the high flux in the year 2000 we projected a launching radius
$> 10^{10}$ cm and densities $> 5 \times 10^{13}$ cm$^{-3}$, which given the higher 
source luminosity seems consistent. 
  
As shown in Sect.~\ref{adcmodels} luminosities which can heat an ADC have been observed during the
high-state of the source. The luminosities inferred during the observations 
reported here are significantly lower.  This suggests that either there are problems with the static corona approximation
or that there are unobserved photons or other heating mechanisms responsible for the coronal excitation.
We do not see any obvious signs for winds in our data like in the year 2000 observations, however there
still seems to be jet activity in the radio (P. Jonker, priv. comm.). On the other hand, the problem
of insufficient heating power for ADCs as derived from X-rays does not seem to be unique to \cirx1. The emission measures
determined in classical ADC sources such as 4U 1822-37~\citep{cottam2001}, EXO 0748-676~\citep{mario2003}, 
2S 0921-63~\citep{kallman2003} and HerX-1~\citep{mario2005}
are greater than the ADC model can produce based on the observed luminosity. In the case of Her X-1 it has been suggested
that the energy range is indeed much dominated by absorption~\citep{kuster2005, ji2007} and the luminosity
in its low-state may be higher than observed. If this is the case, then the ionization balance we observe
in the low-state of \cirx1 should also be largely dominated by obscuration rather than a true
reduction in source luminosity and as such would suggest a genuine analogy to a Seyfert II. 

All the sources mentioned above have high angles of 
inclination ($85^{\circ} \approxlt i \approxlt 90^{\circ}$) 
which given the strong similarity of the low-state spectral properties in \cirx1 is a confirmation
of the previous X-ray projection of a higher angle of inclination. However, there are two distinct
differences between these sources and \cirx1. One is the fact that we have a history of the source 
of once being
very luminous exhibiting powerful winds, the other one is its jet activity which even in the low-flux state
seems to be prevailing.  
Generally almost two orders of magnitude in luminosity are not easily hidden by absorption and the 
relative constancy of 
$N_H$ in \cirx1 over the years is difficult to reconcile with a simple obscuration paradigm. The fact that
the luminosity changed from a level that seemed sufficient to sustain a static corona in the high-state
to a gross deficiency in the low-state without much change in absorption properties between 
obs.~II and III also leads to questioning whether the assumption that the X-ray luminosity is the 
dominating heating component and/or the assumption of a simple static corona is correct. 

Jet and wind activity indicates that the disk
in \cirx1 may be cooling more effciently than predicted from standard models and additional energy is transferred
into the corona or the jet base.
The connection of compact ADCs and the base of jets has been made recently
in a study by ~\citet{markoff2005}. Though it is well beyond the scope of this paper to go into
details, we have to address the possibility that other mechanisms than X-ray illlumination
can heat the corona. Recent studies clearly emphasize the role of magnetic activity.
 
Relativistic jets are a fundamental aspect of accretion on to
black holes~\citep{fender2004b} and much emphasis in recent years
has been devoted to investigate connections between compact
ADCs and jet activity~\citet{merloni2002, markoff2005, krolik2007}.
The formation of jets may arise from radiation pressure~\citep{lynden1978,
begelman1984} or magnetic fields~\citep{blandford1977} and since
we deal with X-ray luminosities well below $L_{edd}$ we may argue
for the latter as the underlying scenario.
Viscous turbulence in the accretion flow is now understood to be
due to the magnetorotational instability (MRI, \citealt{balbus1998}) and it may as well
also the source for a magnetically dominated corona~\citep{merloni2006}.
Observational evidence for MRI effects in GROJ1650-55 has recently been 
presented by~\citet{miller2006c}. The low-flux observations presented
in this paper fit most criteria that qualifies the low-flux state as some form
of low/hard state as seen in accretion black hole binaries (see 
~\citealt{fender2004b, miller2006b} and references therein). 

\subsection{The Absorber Region~\label{photoabsorbers}}

The fact that there is a significant warm absorber in the post-zero phase spectra is obvious through
the broad absorption feature around 1.9~\AA. In hot gases, where the electrons are generally stripped
out of their atomic bond, absorbing fractions are usually generated by the formation of a significant
number of He-like and H-like ion states. Such absorption has been observed in many accreting X-ray
binaries, some show ionization fractions only for iron, some throughout the X-ray band down to oxygen. \circinus
in this respect was pioneering as the \chandra first observations at the end of the high flux period 
showed blueshifted absorption lines for most detectable H- and He-like ions species. On top of this
phenomenon we could show in SBII that the equivalent widths of the hot absorption is tightly 
linked to the slope variation of the underlying continuum.

The changes we see in the warm absorber are a bit more subtle but point in a similar direction.
We now observe an entirely different ionization regime and the atomic physics
is more complex due to the fact that in this absorber we do not have simple He-like and H-like ions.
Atomic data covering this ion regime has not been available until recently when~\citet{kallman2004} updated
the XSTAR database specifically in the Fe K band. Model simulations indicated that under very specific
circumstances 1s-np autoionizing transitions become dominant. The results are Fe XVII - XIX absorption
features near 1.9~\AA\ and the Fe K edge. The appearance of this edge becomes increasingly diluted as
the ionization parameter increases. The model regime then predicts warm columns of 10$^{23.5}$ cm$^{-2}$ and ionization
parameters well below 100. 

The unique ionization regime that produces the observed absorption makes the interpretation of the absorption 
relatively simple and straight forward. An optically thick cool and distant plasma is illuminated by
a moderate source luminosity producing low-ionization ions. What makes the situation more complex is the
fact that we observe at least four absorber regimes: a cold absorber, a ``luke-warm'' absorber during
quiescence, a warm absorber during flares. and last but not least a hot absorber during both phases.  

{\bf The cold absorber:} In all our observations we detect cold photoelectric absorption at levels
significantly above the expected line-of sight-columns. In obs.~III we observe 
about 8$\times10^{22}$ cm$^{-2}$,
in obs.~IV it is lightly lower in quiescence. Compared with obs.~I and II from the 
high flux state of the 
year 2000 these numbers are not so much different. If we compare our results from the partial
coverage fits with the ones in SBII, but also with the ones from the \asca spectra a few years
earlier~\citep{brandt1996} then cold absorption remains at a rather steadly level for long time
periods. The cold absorber should be devoid of significant ionization fractions above Fe X in the 
X-ray band which limits the ioniziation parameter to below 1 and puts the material at a distance
consistent with the outer accretion disk, i.e beyond $10^{11}$ cm. The amount of cold material so
far seems to be fairly invariable. Not only is it rather constant at zero orbital phases, but the edge optical
depth remained constant throughout other binary orbit phases \citep{galloway2007} indicating similar
amounts of cold material at all times.

{\bf The luke-warm  absorber:} The fits at quiescence shows an enormous warm column of
8$\times10^{23}$ cm$^{-2}$. This warm absorber shows up in addition to the cold column. The slight
reduction in cold column can be easily seen at the expense of a slightly increased
source luminosity. It is evident, however, that this large mount of material has not been in the line of sight
during the pre-zero phase obs.~III as otherwise it could have only been exposed through
a luminosity decrease, which would have rendered the warm absorber to a cold one.
There are two possible time lines of how the matter got there. 
Either the material crept up over the time period between the observations, or 
we see newly accreted matter just after the incoming neutron star-disk system re-connected with the companion.
Given the invariability of the cold column over several observing periods and that
a sudden increase in mass accretion at zero orbital phase is expected~\citep{tauris1999} we consider the 
latter possibility as more likely. During the high flux phase we did not see this effect as
there the material was almost completely ionized at all times. This material, on the other hand, is then
likely related to the material causing heavy dipping activity during that flux period~\citet{brandt1996, shirey1999}.

{\bf The warm absorber:} The reason why we separate the absorber in quiescence and during flaring 
periods is that due to the large increase in source luminosity the absorber also experiences a change
in ionization balance which also create higher transition Fe ions. 
The amount of warm column is now only 6.3$\times10^{22}$ cm$^{-2}$ which
is an order of magnitude lower than during quiescence. There may two causes we can consider for the reduction.
First, the source is actually accreting and thus mass moves towards the central source. This flow
is usually subsonic ~\citep{frank1992} and matter moves out of the ionization regime on timescales
much larger than the entire observing time. The light curves in Figure~\ref{lightcurves} feature
flaring periods of only a few hours. More likely is a feedback effect from
direct accretion. Accreted matter is converted into X-rays further heating the absorber.
The increase in cold absorption during flaring may simply reflect the change in ionization state
of accreted matter. 

{\bf The hot absorber:} The fact that we also detect H- and He-like absorbers in the spectra also
indicates the existence of a hot absorption component which reveals itself in resonance absorption
of Ca XX, Fe XXV, and XXVI. The positions of the lines are not blue-shifted as observed during the
high flux phases indicating that this absorber is not a wind but a stationary component. The hot absorber 
is likely part of the warm absorber reflecting  the upper end of the spectrum of ionization fractions.

The light curve of obs.~IV shows that final accretion has two components. In quiescence matter is supplied
at a fairly steady level. During flares accretion also comes in chunks of different sizes which is 
reflected by the very short flare onsets and types. Average rise times are less than a few minutes. Large flares
are accompanied by a staccato of many subflares indicating rather turbulent accretion. Most remarkable is
the correlated change of absorber column and ionization fraction with the luminosity. 
Although our data only allow us to analyze the integrated spectra of quiescent and flaring intervals, 
this analysis indicates that all the absorbers are connected and the source behaviour
may be explained by varying illumination levels of the absorbers combined
with varying obscuration levels of the central source.

There are many places in the system where this can happen. Based on the observed ionization fractions,
most of the material should be at larger radii from the source than the photo-emission regions. This is 
likely also true for the hot absorber, which can be seen by the fact that during flares, the flux of most
emission lines are reduced. The increase in source luminosity cannot remove these ionization fractions
which should survive at even much higher luminosities as we have observed during obs. II (SBII). Although
we cannot rule out other more exotic options a most logical place for the absorbers is the mid-and outer disk
atmosphere/corona which in turbulent systems such as \cirx1 can puff up to fairly large scale heights. 
Even in this case, though, we would limit inclination to higher than 60$^{\circ}$ (see also ~\citet{mario2002})
as we proposed perviously (BSI, SBII, \citealt{schulz2006}). 
The variable absorber activity we observe together with the complex play of warm column and
luminosity (see Table~4) again underlines the importance of obscuration effects and the Seyfert II analogy.

\acknowledgments
We thank all the members of the \chandra\ team for their enormous efforts.
We gratefully acknowledge the financial support of CXC grant GO0-1041X (WNB, NSS) and
Smithsonian Astrophysical Observatory contract SV1-61010 for the CXC (NSS).

\end{document}